\documentclass[fleqn,usenatbib]{mnras}

\usepackage[T1]{fontenc}

\DeclareRobustCommand{\VAN}[3]{#2}
\let\VANthebibliography\thebibliography
\def\thebibliography{\DeclareRobustCommand{\VAN}[3]{##3}\VANthebibliography}

\usepackage{graphicx}	
\usepackage{amsmath}	
\usepackage{amssymb}	
\usepackage{lscape}
\usepackage{ae,aecompl,multirow}
\usepackage{soul}
\usepackage{multirow}
\usepackage{array}

\title[Probing IC/CMB with VHE observations]{Probing IC/CMB Interpretation for the X-ray knots of AGN through VHE observations}

\author[Amal A. Rahman et al.]{
Amal A. Rahman$^{1}$\thanks{E-mail: amalar.amal@gmail.com},
S. Sahayanathan$^{2,3}$\thanks{E-mail:sunder@barc.gov.in},
Malik Zahoor$^{4}$
and P. A Subha$^{1}$
\\
$^{1}$Department of Physics,Farook College,  Calicut University, Kerala-673632, India\\
$^{2}$Astrophysical Sciences Division, Bhabha Atomic Research Centre, Mumbai - 400085, India\\
$^{3}$Homi Bhabha National Institute, Mumbai 400094, India\\
$^{4}$Department of Physics, University of Kashmir, Srinagar 190006, India\\
}

\date{Accepted XXX. Received YYY; in original form ZZZ}

\pubyear{2022}

\begin{document}
\label{firstpage}
\pagerange{\pageref{firstpage}--\pageref{lastpage}}
\maketitle

\begin{abstract}

Detection of hard X-ray spectrum (spectral index $<2$) from the kilo-parsec scale jet of active galactic nuclei cannot be accounted to the synchrotron emission mechanism from the electron distribution responsible for the radio/optical emission. Alternate explanations are the inverse Compton scattering of cosmic microwave background photons (IC/CMB) or synchrotron emission from a second electron population. When the X-ray emission is interpreted as IC/CMB process, the Compton spectrum often peaks at GeV energy and many sources were predicted to be the Fermi candidate sources. The absence of significant gamma ray flux from some of these galaxies by Fermi disfavored the IC/CMB interpretation of the high energy emission. We extend this study to predict the very high energy (VHE) gamma ray emission due to IC/CMB model which can be investigated by Cherenkov Telescope Array Observatory (CTAO). The model parameters deciding the broadband spectral energy distribution are estimated using analytical approximation of the emissivity functions. The emission model is extrapolated to VHE energy and then compared with the CTAO sensitivity. For this particular study, we have selected 18 knots with harder X-ray spectrum and for which the IC/CMB model for X-ray emission was suggested.  

\end{abstract}




\section{Introduction}\label{1}


Localized brightness enhancements found in the  kpc/Mpc scale jets of active galactic nuclei (AGN) are 
commonly termed as knots \citep{2006ARA&A..44..463H}. These knots are well resolved in 
radio and optical wavebands. With the advent of high spatial resolution \emph{Chandra} Telescope, the 
knots were resolved in X-ray bands too. This multi-wavelength emission from AGN knots are generally modeled 
using synchrotron and inverse Compton emission mechanisms. The radio-to-optical emission is well established 
as synchrotron emission from a relativistic electron distribution losing its energy in the jet magnetic
field. On the other hand, the X-ray emission process is either modelled as the high energy tail of 
synchrotron spectrum \citep{2001MNRAS.326.1499H, 2002ApJ...571..206S, 2008MNRAS.388L..49S} or as inverse 
Compton scattering of soft photon field \citep{2000ApJ...544L..23T, 2000ApJ...540L..69S, 2003ApJ...588L..77S, 2011ApJ...739...65P}.

The high energy emission from knots, when interpreted as the inverse Compton mechanism, the target photon field 
at these length scales can be either synchrotron photons \citep{2000ApJ...540L..69S}; usually called as synchrotron self Compton(SSC) 
or the ambient cosmic microwave background radiation \citep{2000ApJ...544L..23T, 2001ApJ...549L.161S}; termed IC/CMB mechanism. The SSC interpretation 
is disfavored as it demands a magnetic field that largely deviates from equipartition \citep{2000ApJ...544L..23T, 2006ARA&A..44..463H}. 
The IC/CMB model; however, requires the jets to be relativistic even at kpc scales so that in the knot frame, 
the ambient photon field is relativistically boosted to overpower the synchrotron photon energy 
density. Nevertheless, this interpretation is capable of reproducing the observed X-ray flux from the 
knots and satisfies near equipartition magnetic field \citep{2003ApJ...588L..77S, 2000ApJ...540L..69S, 2000ApJ...544L..23T}.

The X-ray flux from the knots, when interpreted as IC/CMB process, the Compton spectrum peaks at GeV energy
with significant radiation at gamma-ray energies. Hence, \emph{Fermi} observations play a crucial role to validate this 
emission scenario \citep{2006ApJ...653L...5G}. 
However, even with a decade of observations, \emph{Fermi} failed to detect any appreciable
$\gamma$-ray photons from some sources and the upper limits drawn fall significantly below the flux predicted 
by the IC/CMB emission model \citep{2014ApJ...780L..27M,2015ApJ...805..154M}. 
Particularly, the $\gamma$-ray studies of the \emph{Chandra} detected  sources namely,{ 3C\,273(CDQ), PKS\,0637-752(CDQ), PKS\,1136-135(LDQ), PKS\,1229-021(CDQ), PKS\,1354+195(CDQ) and PKS\,2209+080(CDQ)} \citep{2014ApJ...780L..27M,2015ApJ...805..154M, 2017ApJ...849...95B} have 
strongly ruled out the IC/CMB interpretation but favour a scenario where synchrotron emission 
from a second population of relativistic electrons being responsible for the observed X-ray 
emission \citep{2015ApJ...806..188L, 2021MNRAS.501.6199T, 2002ApJ...565..244H,2004ApJ...613..151A, 2005ApJ...622..797K}. In our recent work, we show that the 
advection of electrons from the regions of particle acceleration can naturally 
produce two distinct populations which can successfully explain
the X-ray emission from the kpc scale knots of 3C\,273 \citep{2022MNRAS.515.1410R}.

Though recent observational evidence, including $\gamma$-ray upper limits, disfavour the IC/CMB interpretation
of X-ray emission \citep{2018PhDT.......153B}, it is still a preferred model for those sources which are not yet 
ruled out by \emph{Fermi} observations \citep{2012ApJ...748...81K, 2012ApJ...755..174G, 2015ApJ...807...48S}. \cite{2010ApJ...710.1017Z} present the analysis of 22 hotspots and 
45 knots to conclude that IC/CMB model can explain the X-ray emission from the majority of the jet components.

This study was then extended to a larger number of jet components and the jet power is estimated under IC/CMB and 
SSC emission scenarios for the observed X-ray fluxes \citep{2018ApJ...858...27Z}. The jet power estimated
considering IC/CMB strongly correlate with the jet kinetic power obtained through radio studies. Whereas,
the SSC interpretation did not exhibit any significant correlation. These findings further supported the 

IC/CMB as a promising mechanism to explain the high energy emission from the knots.

{\cite{2017MNRAS.466.4299L} revisited some aspects of the IC/CMB model to show the role of electron cooling in shaping the spectrum. They suggest, the overproduction of gamma-rays can be avoided by suppressing the high energy end of the emitting particle population. They applied this model for the case of PKS 0637-752 and demonstrated that IC/CMB can still be a valid explanation for the high energy production from the large scale knots of AGN. In this context, it will be useful to explore the alternative techniques which can supplement the Fermi studies to test the plausiblity of the IC/CMB interpretation for the high energy emission.}

Modelling the X-ray emission from kilo-parsec scale jets using IC/CMB model also suggests this spectrum
can extend up to very high energies (VHE)\citep{2021AAS...23723805M, 2019ApJ...883L...2M, 2006ApJ...653L...5G}. Hence, future VHE observation of these sources can provide further
constraints in addition to the ones drawn through \emph{Fermi} observations. The operational ground-based VHE telescopes
employing imaging atmospheric Cherenkov techniques have already detected six radio galaxies with mis-aligned jets 
at GeV/TeV energies \citep{2012A&A...539L...2A, 2009ApJ...695L..40A, 2006Sci...314.1424A, 2011ICRC...12..137H, 2015ICRC...34..801D, 2010ApJ...723L.207A, 2022Galax..10...61R}. Among these sources four belong to FRI class of radio 
galaxies (M\,87, Cen\,A, NGC\,1275, and 3C\,264) while other two (PKS\,0625-35 and IC\,310) show the properties of radio 
galaxy and blazar as well \citep{2008ICRC....3.1341W}. If IC/CMB is a viable emission process for the X-ray knots,
then this suggests many misaligned jets may be a potential TeV candidate. Alternatively, VHE studies can also be a tool
to validate this emission scenario in tandem with the \emph{Fermi} observations. 

Considerable advancement in the stereoscopic imaging atmospheric Cherenkov techniques have led to new generation
VHE telescopes whose sensitivity are appreciable even at few tens of GeV. Particularly, with the inputs from the 
upcoming Cherenkov Telescope Array (CTA), it may be possible to verify the IC/CMB interpretation of the X-ray knots 
which were not been ruled out through \emph{Fermi} observations. In the present work, we selected all the kilo parsec
scale X-ray jets for which IC/CMB emission process is favoured. We model the radio and optical fluxes from these sources 
as synchrotron emission from a broken power-law electron distribution. The source parameters are constrained 
considering equipartition between the magnetic field and the particle energy density. The IC/CMB spectrum is 
then extrapolated to VHE energies and compared with the CTA sensitivity curves. The paper is organised as follows:
In section 2, we describe the source selection and in section 3, the spectral models used in the paper. In section 4 we present our 
results and discussion. Throughout this work 
we consider a cosmology where 
$H_0 =71$ km s$^{-1}$ Mpc$^{-1}$, $\Omega_m = 0.27$ and $\Omega_\Lambda = 0.73$.

\begin{table*}

  \centering
\begin{tabular}{|lccccccc|}
\hline
\multirow {1}{*}  Source & Type & Knot & $z^a$ & $Log({F}_{\rm obs}^{\rm r})^b$ & $Log({F}_{\rm obs}^{\rm o})^c$ & $Log({F}_{\rm obs}^{\rm x})^d$ & Reference \\ \hline
3C\,15 & FR\,I  & K\,C & $0.073$  & $-24.63$ & $-28.85$ &  ${-31.87^{\rm +0.119}_{\rm -0.108}}$ & { \citep{2007MNRAS.374.1216D,2003A&A...410..833K}}   \\ 
3C\,17 & hybrid & S\,3.7 & $0.22$ & $-24.51$ & ${-28.66}$ &  ${-32.19^{\rm +0.256}_{\rm -0.168}}$ & \cite{2009ApJ...696..980M}\\ 
3C\,17 & hybrid & S\,11.3 & $0.22$ & $-24.10^{\rm +0.04}_{\rm -0.04}$ & ${-29.64}^{\rm +0.07}_{\rm -0.07}$ &  ${-32.27^{\rm +0.15}_{\rm -0.15}}$ & \cite{2009ApJ...696..980M}\\ 
3C\,111 & FR\,II & K\,22 &   $0.049$ & $-25.42^{\rm +0.05}_{\rm -0.06}$ & $-29.38^{\rm +0.19}_{\rm -0.18}$ &  $-32.40^{\rm +0.12}_{\rm -0.12}$ & \cite{2016ApJ...826..109C} \\
3C\,111 & FR\,II & K\,30 &   $0.049$ & $-24.86$ & $-28.04$ &  ${-31.13}$ & \cite{2016ApJ...826..109C} \\
3C\,111 & FR\,II & K\,61 &   $0.049$  & $-24.43$ & ${-28.78}^{\rm +0.12}_{\rm -0.11}$ &  ${-31.38}$ & \cite{2016ApJ...826..109C}\\ 
3C\,120 & {FR\,I} & K\,4 &   $0.033$  & $-24.45$ &  $-28.73$    & $-31.52^{\rm +0.06}_{\rm -0.06}$ &\cite{2004ApJ...615..161H}\\
3C\,120 & {FR\,I} & S\,2 &  $0.033$   & $-24.88^{\rm +0.05}_{\rm -0.05}$ &  $-28.62^{\rm +0.06}_{\rm -0.06}$    & $-32.54^{\rm +0.17}_{\rm -0.17}$ &\cite{2004ApJ...615..161H}\\
PKS\,1354+195 & CDQ & S\,4.0 &   $0.720$ & $-24.96$ & $-30.07^{\rm +0.09}_{\rm -0.09}$ &  ${-32.27}^{\rm +0.06}_{\rm -0.06}$ &\cite{2017ApJ...846..119H} \\
PKS\,1354+195 & CDQ & S\,5.3 &   $0.720$ & $-25.54^{\rm +0.04}_{\rm -0.04}$ & $-30.23^{\rm +0.09}_{\rm -0.09}$ &  ${-32.78}^{\rm +0.11}_{\rm -0.11}$ &\cite{2017ApJ...846..119H} \\
3C\,346 & FR\,I & K\,C &   $0.161$   & $-24.16$ & $-28.40^{\rm +0.13}_{\rm -0.13}$ &  ${-31.81}^{\rm +0.05}_{\rm -0.06}$ &\cite{2005MNRAS.360..926W} \\
3C\,454.3 & CDQ & K\,A &   $0.859$ & $-24.39$ & $-29.92^{\rm +0.09}_{\rm -0.10}$ &  ${-31.24}^{\rm +0.08}_{\rm -0.11}$ &\cite{2007ApJ...662..900T} \\
3C\,454.3 & CDQ & K\,B &   $0.859$  & $-23.71^{\rm +0.04}_{\rm -0.04}$ & $-29.31^{\rm +0.03}_{\rm -0.03}$ &  ${-31.25}^{\rm +0.11}_{\rm -0.10}$ &\cite{2007ApJ...662..900T} \\
PKS\,2101-490 & CDQ & K\,6 &   $1.040$ & $-25.12^{\rm +0.02}_{\rm -0.02}$ & $-29.89$ &  ${-32.}^{\rm +0.11}_{\rm -0.11}$ &\cite{2012ApJ...755..174G} \\
PKS\,B0106+013 & CDQ & K\,1 &   $2.11$ & $-24.86$ & $-30.50$ &  ${-33.08}$ &\cite{2012ApJ...748...81K} \\
PKS\,B0106+013 & CDQ & K\,2 &   $2.11$   & $-24.62$ & $-30.41$ &  ${-33.37}$ &\cite{2012ApJ...748...81K} \\
PKS\,B0106+013 & CDQ & K\,3 &   $2.11$   & $-23.74$ & $-30.62$ &  ${-33.53}$ &\cite{2012ApJ...748...81K} \\
PKS\,1045-188 & CDQ & K\,C &   $0.590$  & $-24.62^{\rm +0.09}_{\rm -0.09}$ & $-30.09^{\rm +0.05}_{\rm -0.05}$ &  ${-32.11}$ &\cite{2015ApJ...807...48S} \\

\hline
\end{tabular}\caption{List of AGNs with Jet Knots/components included in our source list.
{Notes:} $^a z$, redshift;{$^b$ (col V), $^c$ (col VI), $^d$ (col VII) are the average observed fluxes in radio, optical and X-ray energies respectively. The energy range over which this fluxes are averaged is mentioned in references (col VIII).}} 

\label{tab3}
\end{table*}
\section{Source selection}
\emph{Chandra} during its two decades operation was able to resolve nearly 150 X-ray extended features from 
AGN\footnote{Xjet:https://hea-www.harvard.edu/XJET/}. Many of these extended jet features
showed bright knots and hotspots which are coincident (near-coincident) with radio/optical maxima. Spectral
energy distribution (SED) of these knots (and unresolved jet features) in radio, optical and X-ray have been
already analysed and reported in earlier works \citep{2006ApJ...647L.107S, 2007ApJ...657..145S, 2001ApJ...547..740W, 2000ApJ...542..655C}. Due to less number of counts, spectral resolution at these
energies cannot be achieved and the fluxes were reported only for one or few energy bins \citep{2011ApJS..197...24M, 2018ApJS..234....7M, 2018ApJS..235...32S}. However, 
convoluting the X-ray instrumental response with a power-law source spectrum, constraints on the spectral index
can be obtained and this can be helpful in identifying the emission process.

{Among the sample of extended X-ray jets collected from the literature, we selected 18 X-ray knots/components with radio-optical-X-ray observations and a harder X-ray spectrum (spectral index less than 2 or the X-ray flux value larger than high-energy extension of the radio-to-optical synchrotron spectrum). The X-ray emission from these knots were interpreted as IC/CMB. Additionally, for these sources, IC/CMB model were not ruled out by Fermi studies until 2023. In Table \ref{tab3} we provide the list of complete sources and knots.}

\section{IC/CMB model}

The radio/optical/X-ray knot buried in AGN jet is considered to be an spherical region of size $R_{\rm knot}$ governed by its 
radio contour. We assume this region is uniformly populated with a broken power-law electron distribution of
the form given by
\begin{align}\label{eq:bpl}
  N(\gamma)d\gamma=\begin{cases}
               K\gamma^{-p}d\gamma  \hspace{15mm}    \gamma_{\rm min} < \gamma < \gamma_{b}\\
              K\gamma_b^{q-p}\gamma^{-q}d\gamma  \hspace{10mm}  \gamma_{b} < \gamma < \gamma_{\rm max} 
            \end{cases}
\end{align}
where, $K$ is the normalisation, $\gamma$ is the Lorentz factor of the electron {and $\gamma_b$ is the radiative cooling break}. 
The electrons lose their energy through synchrotron 
and inverse Compton radiative processes. If we consider $P_{rad}(\gamma,\nu)$ as the single particle 
emissivity due to these radiative processes, then the total emissivity from the knot can be obtained as \citep{1986rpa..book.....R}
\begin{align}
	J_{\rm rad}(\nu)= \frac{1}{4\pi}\int\limits_{\gamma_{\rm min}}^{\gamma_{\rm max}}P_{\rm rad}(\gamma,\nu)N(\gamma)d\gamma     
\end{align}         
Here, $P_{\rm rad}$ additionally depend on the magnetic field in case of synchrotron (rad $\rightarrow$ syn) 
while in case of inverse Compton process, it depends on the target photon energy and distribution (rad $\rightarrow$ ic).

We assume a tangled magnetic field configuration which is in equipartition with the electron distribution {(condition that indicates total energy of the system is minimum \citep{1959ApJ...129..849B})};
\begin{align}\label{eq:beq}
	\frac{B^2}{8\pi} = m_e c^2 \int\limits_{\gamma_{\rm min}}^{\gamma_{\rm max}} N(\gamma) \gamma d\gamma
\end{align}
Under delta function approximation for single particle emissivity, an approximate emissivity for synchrotron 
emission can be obtained as \citep{2009herb.book.....D,2012MNRAS.419.1660S}
\begin{align}\label{eq:flux1}
	J_{\rm syn}(\nu) \approx \frac{c\sigma_{T}B^2}{48\pi^2} \nu_L^{\frac{-3}{2}} N\left(\sqrt{\frac{\nu}{\nu_L}}\right) \nu^{\frac{1}{2}}
\end{align}
Since the X-ray emission is interpreted as IC/CMB process, we can arrive an analytical form of the IC emissivity
by considering the CMB distribution as a monochromatic photon field and a delta function approximation for the 
$P_{\rm ic}$ 
\citep{2009herb.book.....D,2018RAA....18...35S}
\begin{align}\label{eq:flux2}
	J_{\rm ic}(\nu) \approx \frac{c\sigma_{T}U_*}{8\pi{\nu}_*} \sqrt[]{\frac{\Gamma\nu (1+\mu)}{{\nu}_*}} N\left[\sqrt{\frac{\nu}{\Gamma {{\nu}_*}(1+\mu)}}\right]
\end{align}
Here, $\nu_*$ and $U_*$ are the frequency and energy density of the external photon field, $\Gamma$ is the bulk
Lorentz factor of the jet and $\mu$ is the cosine of the jet viewing angle measured in the proper frame of the AGN. 
In the observer's frame, the total flux due to these radiative processes can be obtained after considering the 
relativistic and cosmological effects \citep{1984RvMP...56..255B, 1995ApJ...446L..63D}
\begin{align}\label{eq:obs_flux}
	F_{\rm obs}(\nu_{\rm obs})= \frac{\delta_D^3(1+z)}{d_L^2}VJ_{\rm syn/ic}\left(\frac{1+z}{\delta_D}\nu_{\rm obs}\right)
\end{align}
Here, $V = 4/3\, \pi R_{\rm knot}^3 $ is the volume of the emission region, $z$ is the source redshift, $d_L$ is the luminuousity distance 
and $\delta_D = 1/[\Gamma(1-\beta \mu)]$ is the jet Doppler factor.

\subsection{Source Parameters}

The observed spectrum due to IC/CMB model is mainly governed by 8 source parameters with 4 of them 
$K$, $p$, $q$ and $\gamma_b$ governing the electron distribution, and the rest are the size of the emission 
region $R_{\rm knot}$, 
magnetic field $B$, jet Lorentz factor $\Gamma$ and the jet viewing angle $\theta$. Besides these, the parameters
$\gamma_{\rm min}$ and $\gamma_{\rm max}$ do not govern the source fluxes rather decide the low and high energy
end of radiation spectrum (see section \ref{sec:4}). Limited amount of information available in radio, optical and X-ray bands do not let us 
to draw stringent constraints on these parameters. However, modest constraints can be imposed by assuming 
equipartition between the magnetic field and particle energy densities (equation \ref{eq:beq}) and
the emission region size of kilo-parsec scale order which is consistent with the typical radio contours of the knots.

The synchrotron flux at radio/optical in combination with equation \ref{eq:beq} can effectively constrain $B$
and $K$ while, the spectral indices can be used to identify $p$ and $q$. 
The synchrotron spectral peak frequency can be expressed in terms of the source parameters as
\begin{align}
	\nu_{\rm p,syn} = \frac{\delta_D}{1+z} \gamma_b^2 \frac{eB}{2\pi m_e c}
\end{align}
For a given $\delta_D$, $\gamma_b$ can be estimated from the approximate peak frequency. With these constraints 
we will be left with the parameters $\Gamma$ and $\theta$ which can be fine-tuned to reproduce the X-ray flux due to
IC/CMB process. The approximate analytical expressions for the synchrotron and IC/CMB fluxes are used only to 
estimate the source parameters. The final model spectrum is produced numerically using exact functional form of the 
single particle emissivities and compared with the observed fluxes.

{An upper limit on the angle of the jet to the line of sight ($\theta$) can be drawn from the apparent superluminal motion obtained through Very Long Baseline Interferometry (VLBI) observations. \cite{2023MNRAS.518.3222B} estimate the upper limits on $\theta$ for 3C\,111, 3C\,120, PKS\,1045-188 and 3C\,454.3 to be $13.5^0, 12.6^0, 10.5^0$ and $4.5^0$ respectively. An upper limit of $13^0$ is estimated for jet inclination angle of PKS\,B0106+013 \citep{2012ApJ...748...81K}. \cite{2017ApJ...846..119H} using the apparent superluminal proper motions of pc-scale jet estimate that the kpc-scale jet of PKS\, 1354+195 is aligned at $\theta \leq 12^0$. \cite{2009MNRAS.398.1207D} calculate the viewing angle of 3C 346 to be $\theta = 14\pm8^0$. The apparent velocity measurements of 3C 15 is not available. However, \cite{1997MNRAS.291...20L} found its jet/counter-jet flux density ratio and puts a constraint of $\theta \approx 45^0-50^0$ for $\beta \geq 0.9$. But, in order to explain the X-ray emission through IC/CMB model, the jet inclination should be less than the angle constraint drawn from jet/counter-jet ratio. Therefore in this work, $\theta$ is chosen such that the model curve fits with the observations. The apparent velocity or jet/counter-jet measurements for sources 3C\,17 and PKS\,2101-490 are not available. Therefore, viewing angle for these sources are chosen such that the IC/CMB model curve can fit with the radio-optical-X-ray observations of the knots.
The sizes of individual knots are constrained using radio data and the value of $R_{size}$ is listed in table 2.}

\section{Results and Discussion}\label{sec:4}

We apply the IC/CMB model to reproduce the X-ray flux of 18 knots/jet components from the sources which are not ruled 
out by initial \emph{Fermi} observations (Table \ref{tab3}). As mentioned earlier, the initial guess parameters are derived from approximate 
analytical expressions for the emissivity functions and are then fine tuned to reproduce the radio-optical-X-ray fluxes
using exact numerical solution. The best fit model parameters (chi-by-eye) for these knots are given in Table \ref{tab2}
with their corresponding SEDs shown in Figure 1 and 2.                                             

{The parameters $\gamma_{min}$ and $\gamma_{max}$ cannot be constrained from the available information and we have fixed these quantities at $50$ and $10^3 \times \gamma_b$ ($\sim 10^8-10^9$) resectively for all the SEDs. The Compton up-scattered photon energy scales as $\gamma^2$ and  hence this choice of $\gamma_{max}$ assures the Compton spectral component extend up to sixth order from the spectral peak. The gyro-radius of the electron corresponding to $\gamma_{max}$ is $\sim 10^{-2}- 10^{-1}$ pc, which is much less than the assumed size of the knot. Therefore, this ensures that such high energy electrons can still be confined within the emission region. The equipartition magnetic field depends on $\gamma_{min}$, since the particle energy density is largely decided by the electrons at this energy. However, we noted that moderate variation in $\gamma_{min}$ (within the same order) do not alter our conclusions significantly.}

In order to predict the VHE flux from these knots/jets, we extended the IC/CMB model spectrum to VHE energies. The VHE photons 
produced from distant sources undergo significant attenuation through pair production process with the extragalalctic background 
light (EBL). To account for this, we have considered the EBL spectrum provided by \cite{2017A&A...603A..34F} and the {attenuated VHE
spectrum is compared with the 50 hour sensitivity curve of CTAO obtained from CTAO webpage\footnote{https://www.cta-observatory.org/science/cta-performance/}}. In Fig 1 and 2, the blue
solid line is the VHE spectrum after accounting for the EBL induced attenuation. The red, green and magenta solid line are the 50 hour CTAO differential sensitivities of Omega, 
Alpha (both Northern array) and Omega configuration (Southern array) respectively.
The \emph{Fermi} gamma-ray upper limits/detections are shown as green color inverted triangles/solid circles.
Out of all the knots studied, we find the knots of 3C\,111 (K\,30), 3C\,120 (K\,4) and a section of 3C\,120 jet (S\,2) fall within the 
detection threshold of CTAO. The SED corresponding to these knots/jet is shown in Fig. 1 along with the CTAO sensitivity curve.
The knots whose VHE model flux fall below the CTAO sensitivity curve are shown in Fig. 2. {Moreover, both 3C\,120 and 3C\,111 can be detected from the Northern site of the Cherenkov Telescope Array (CTA) at reasonable zenith angles.}

Among the two sources predicted as a CTAO VHE candidate under IC/CMB model, 3C 111 belongs to FRII morphology \citep{1974MNRAS.167P..31F} 
located at z=0.049. The radio observation suggests a 100 kilo-parsec long jet and \emph{Chandra} was able to resolve
nearly 10 knots with at least eight which are prominent in X-rays \citep{2016ApJ...826..109C}. Among them, the knots  K\,22, K\,30 and K\,61 
indicate a hard X-ray spectrum suggesting an inverse Compton origin of the X-ray emission. Whereas, the knots K\,9, K\,14, K\,38, K\,45, K\,51, 
K\,97 and K\,107 has steeper X-ray spectrum supporting a synchrotron origin \citep{2018ApJ...858...27Z}. 
Hence, for the present work we have considered only the three X-ray knots K\,30, K\,61 and K\,22, with K\,30 being the brightest among them.
Modelling the X-ray emission as IC/CMB mechanism predicts significant VHE flux for the knot K\,30 which can be {examined} by future CTAO 
observations. The predicted VHE flux of K\,61 and K\,22 fall below the CTAO sensitivity though the X-ray flux of K\,61 is comparable with K\,30. 
We anticipate the reason for this being the steep X-ray spectrum of K\,61 as compared to K\,30.

Recently, \citep{2023MNRAS.518.3222B} ruled out 
the IC/CMB interpretation of the X-ray emission for 3C\,111 considering the knot K\,61 through the updated \emph{Fermi} upper limits (these upper limits are shown as blue inverted triangles in Fig. 1 and 2). However, 
the steep X-ray index of the knot K\,61 is not consistent with the model spectrum and this can cause ambiguity (See Fig. A6 in \citep{2023MNRAS.518.3222B}). On 
the other hand, the X-ray 
spectrum of K30 is hard and can provide better constraints. Comparing the predicted gamma ray flux of this knot with the updated \emph{Fermi} 
upper limits again disfavours the IC/CMB interpretation. {Nevertheless, studying this source at VHE will be an additional confirmation to this 
conjecture or may provide stronger constraints on the emission model if detected.} 

The second source predicted as a VHE candidate for CTAO under IC/CMB model is 3C\,120 located at z=0.033 which is a broad line radio galaxy initially classified as Seyfert I \citep{2015A&A...574A..88S}. The radio morphology of the source is similar to FR I class with radio structures extending up to 100 kpc and superluminal components identified with VLBI 
studies \citep{2001ApJ...556..756W,1987ApJ...316..546W}. The X-ray jet consists of four bright knots K4, K7, K25 and K80 and two sections 
of the jet S\,2 and S\,3 bright in X-rays \citep{1999ApJ...518..213H}. The X-ray spectrum for the knot K\,7 and the jet component S\,3 is steep supporting a synchrotron 
origin while K\,4 and S\,2 suggests IC/CMB origin due to hard X-ray spectrum. The knots K\,25 and K\,80 are not well resolved in optical 
waveband \citep{2004ApJ...615..161H}. Hence, in this work we have considered only the knot K\,4 and the jet component S\,2, and modelled their X-ray emission
through IC/CMB process. Extrapolating this emission model to VHE suggests, the gamma-ray flux from these regions fall within the 
50-hour sensitivity of CTAO. In Fig. 1, we show the model spectrum corresponding to synchrotron and IC/CMB emission processes along with 
the observed radio--optical--X-ray fluxes. 
The CTAO sensitivity curve is shown in red solid line while the \emph{Fermi} gamma-ray upper limits/detections are
shown as green color inverted triangles/solid circles. Though the model curve falls within this \emph{Fermi} upper limits, 
recent upper limits(blue inverted traingles in Fig.1) derived from 12 years of observation again disfavours the IC/CMB model for this source also \citep{2023MNRAS.518.3222B}.

The analytical approximation for the inverse Compton emissivity (equation (\ref{eq:flux2})) is arrived considering the scattering 
process to happen in Thomson regime. However, at VHE energy the scattering may be inelastic and better described by the  Klein-Nishina 
cross section. In order to investigate this, we draw limits on the target photon energy beyond which the 
scattering process deviates from the Thomson condition \citep{2009MNRAS.397..985G}. This happens
when the target photon energy measured in the electron rest frame exceeds (or equal to) the rest mass energy \citep{1970RvMP...42..237B}. 
Using this condition along with the scattered photon energy estimated under Thomson regime, one can arrive the condition 
on target photon frequency as \citep{2009MNRAS.397..985G, 2012MNRAS.419.1660S}

\begin{align}\label{eq:thomson}
	\nu^* &\lesssim \frac{1}{\nu(1+z)} \left(\frac{\delta_D}{\Gamma}\right)\left(\frac{mc^2}{h} \right)^2 \nonumber  \\
	&\lesssim 10^{14} \left(\frac{\delta_D}{\Gamma}\right) \left(\frac{\nu}{10^{26}} \right)^{-1} \quad \textrm{Hz}
\end{align}
where, $\nu^*$ is the frequency of target photon in AGN frame, $\nu$ is the scattered photon (VHE) frequency in observer frame and
$\delta_D/\Gamma \approx 1.5$ (for $\Gamma = 4.0$ and $\theta = 9^0$ corresponding to the knot K\,30 of 3C\,111). 
For IC/CMB process the target photon frequency is $\approx 1.6 \times 10^{11}$ Hz and the scattering falls within the Thomson regime.
Nevertheless, in the numerical code used to reproduce the model spectrum we have considered the exact Klein-Nishina cross section 
for the IC scattering \citep{1993ApJ...416..458D}.

{The jet components(or knots) considered in this work are located at kilo-parsec scale distances from the AGN core. For instance, the jet components that fall within the detection threshold of CTAO, K\,4 and S\,2 of 3C\,120 are at distances larger than 2.5 kpc from their core \citep{1999ApJ...518..213H}. And the knot K\,30 of 3C\,111 is approximately at a distance of 60 kpc from the core \citep{2016ApJ...826..109C}. At this large scale distances, the dominant photon field is Cosmic Microwave Background Radiation(CMBR) compared to other photon fields associated with galaxy. Additionally, the relativistic boosting increases the energy density of the CMB in the jet frame by $\Gamma^2$ while, other external photon field like starlight, thermal infrared radiation from the dusty torus and emission from the accretion
disk fall behind the relativic motion of the emission region and hence, their energy density will be reduced by $\approx \Gamma^2$ \citep{1984RvMP...56..255B}. \cite{2011MNRAS.415..133H} has modelled the TeV $\gamma$-ray emission from M\,87 and Cen\,A considering inverse Compton scattering of various photon fields by the high-energy electrons responsible for the synchrotron X-rays on kiloparsec scale. However, for this study they have considered only the inner jet region which fall within a kpc from core and the contribution of photon fields other than CMB are also substantial.}

In this work, we attempt to study the AGN with misaligned jets which are the probable VHE candidates for CTAO. Such misaligned AGN jets have already 
been studied at VHE and the first significant detection was the FR I source M87 located at a distance 16 Mpc (z = 0.0043)\citep{2006Sci...314.1424A}. Observations
of this source from 2003 till 2006 by H.E.S.S witness the flux variability in timescale of few days. Constraints on the emission region size drawn 
from this variability timescale suggests the VHE emission to be arising very close to the central engine. The plausible locations are the bright inner
most knot HST-1 or the nucleus of M\,87. Different emission models have been proposed favouring either of these regions albeit a consensus 
was never arrived \citep{2008A&A...479L...5R, 2008A&A...478..111L, 2011ApJ...730..123L, 2011MNRAS.415..133H, 2012grb..confE.131F, 2012AIPC.1505...80R}.

The other misaligned AGN which are detected in VHE are Cen\,A \citep{2009ApJ...695L..40A},  NGC\,1275 \citep{2012A&A...539L...2A}, 3C\,264 \citep{2011ICRC...12..137H}, PKS\,0625-354 \citep{2018MNRAS.476.4187H} and IC\,310 \citep{2010ApJ...723L.207A}. An important challenge often 
encountered while explaining the VHE emission from these sources is the necessity of substantial Doppler boosting. In case of blazars, 
the jet is aligned close to the line of sight and hence the emission is significantly Doppler boosted.  This allow the gamma ray emission
to overcome the pair production losses since the rest frame photon energy is less than the observed energy \citep{1995MNRAS.273..583D}. 
In case of misaligned AGN, the Doppler boosting will be minimal due to large viewing angle  of the jet. This introduces 
strong constraints on the choice of the bulk Lorentz factor of the jet and the viewing angle \citep{2016ApJ...827...66S}. However, the viewing angle inferred
from studying the broadband spectral energy distribution of the source often conflicts with the ones obtained through radio
morphological studies \citep{2011A&A...531A..30R, 2017MNRAS.465L..94F}. 

The low energy tail of the electron distribution upscattering the CMB photons to X-ray energies are also responsible for the radio 
emission through synchrotron process. This predicts the knots observed at these energy bands to be co-spatial. 
However, the observed offset between X-ray and radio centroids of the knots from the large scale jets contradict 
this co-spatiality \citep{2021ApJS..253...37R}. The relatively longer cooling time of low energy electrons also 
suggest the X-ray emission to be persistent downstream in the jet along with radio. However, in many cases, 
X-ray emission is localized and do not show continuity similar to radio maps \citep{2006ApJ...640..211H}. 
These observational features advocate against the IC/CMB interpretation of the X-ray emission.

Another line of evidence challenging IC/CMB is the detection of X-ray counter jet in 
Pictor A \citep{2016MNRAS.455.3526H}. The jet counter-jet ratio for this source suggests 
the jet viewing angle $\theta <45^0$ and the jet velocity $v \leq 0.5 c$. 
The IC/CMB models for the X-ray emission; on the other hand, demands larger 
jet velocity to boost the CMB photons significantly in the frame of the emission
region. Accordingly, smaller viewing angles are also assumed. 
Similarly, the high degree of X-ray polarisation witnessed in certain knots \citep{2013ApJ...773..186C}, 
the knot flux variability \citep{2016MNRAS.455.3526H} and the deficit of correlation between the ratio 
of X-ray to radio jet luminosity with the source redshift \citep{2021ApJ...914..130S} poses strong challenges to the IC/CMB 
interpretation of the X-ray emission. Observation of the sources 3C\,111 and 3C\,120 in VHE can further verify
this emission model and in addition, provide clues regarding the high energy emission process.

\begin{figure*} \label{fig1}
\includegraphics[angle=0,scale=0.55]{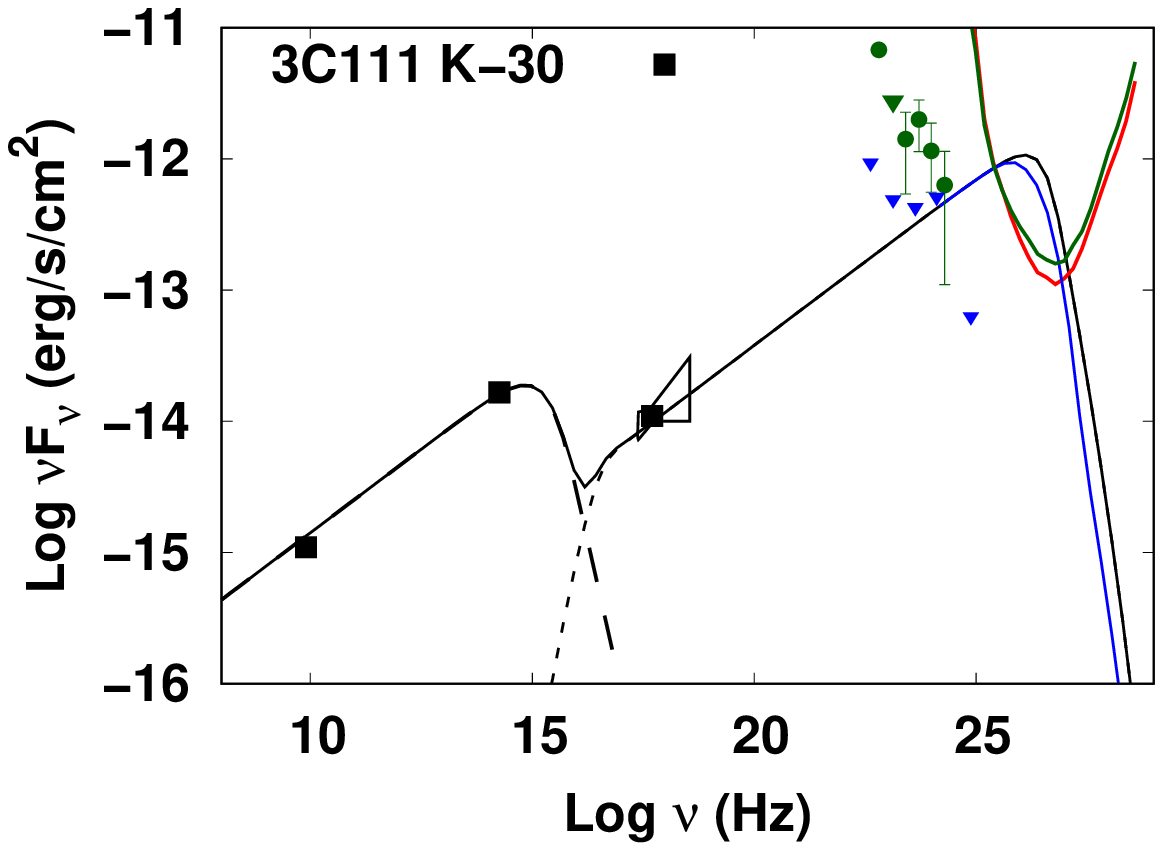}

\includegraphics[angle=0,scale=0.55]{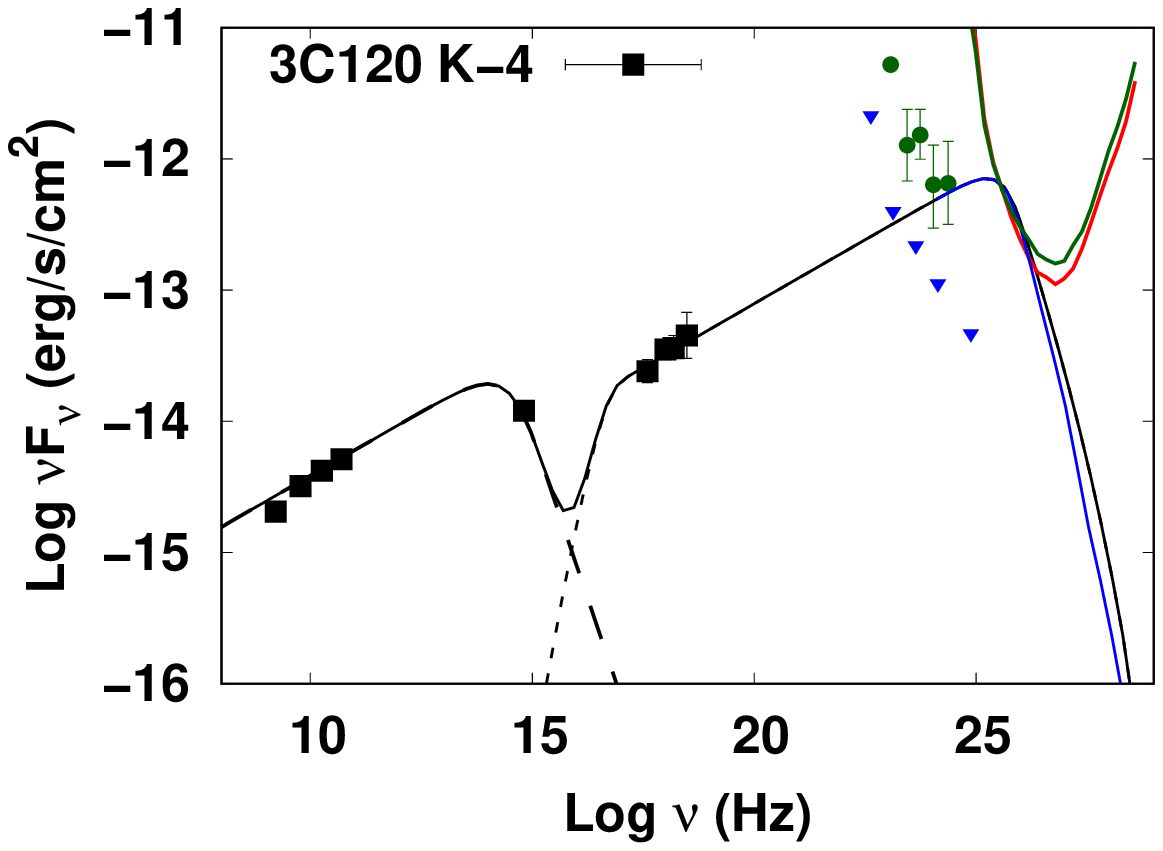}
\includegraphics[angle=0,scale=0.55]{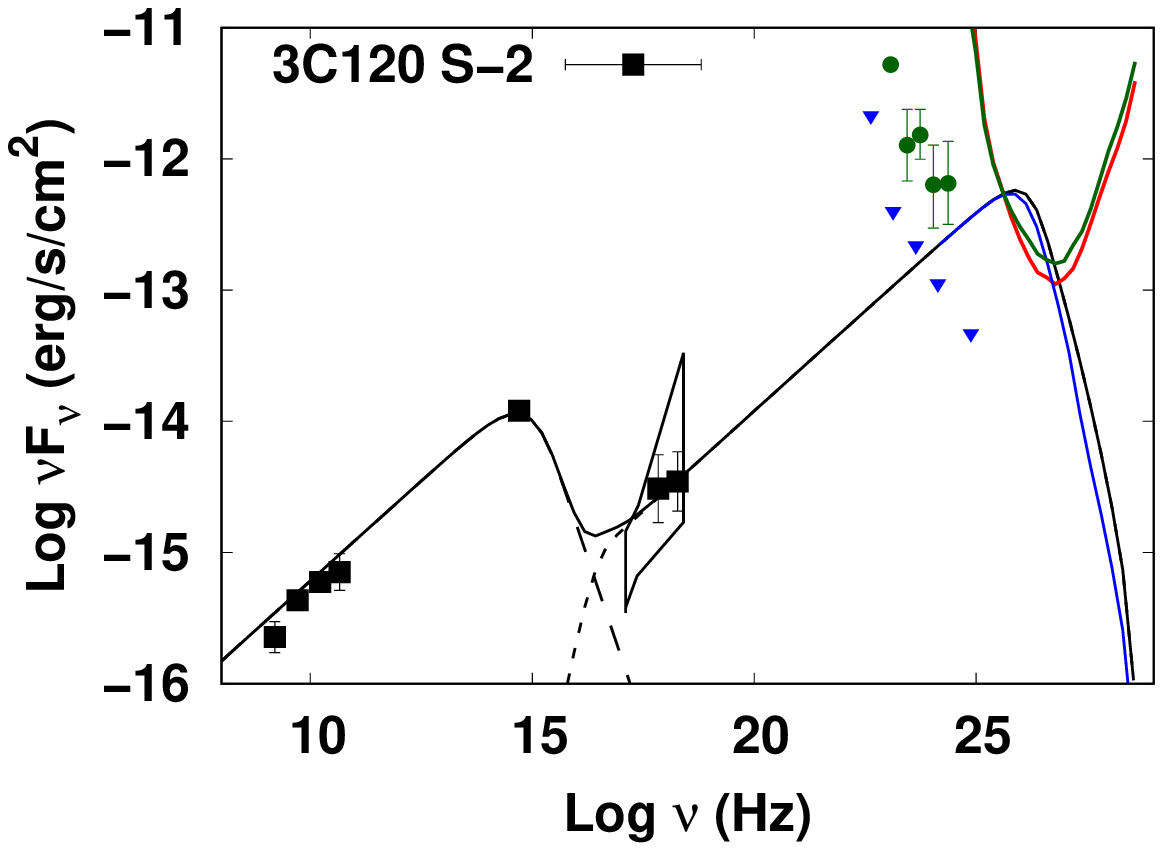}

\caption{Large scale knots falling within the detection threshold of CTAO. Dashed line and densely-dashed line are the synchrotron and IC/CMB model curves respectively. The black solid line represents the synchtron+IC/CMB curve. Solid squares are the multiwavelength observational data of the knots. {Red solid line and green solid line is the differential sensitivities of CTAO-Northern array Omega(50 hour) and Alpha configuration respectively}. Blue solid line is the IC/CMB model curve corrected for EBL absorption.Inverted triangles(green) and solid circles(green) are the \emph{Fermi} upper limit values and \emph{Fermi} observations\citep{2017RAA....17...90X}. Inverted triangles(blue) are the updated \emph{Fermi} upper limit values \citep{2023MNRAS.518.3222B}. {Fermi-LAT points(solid circles) correspond to the total observed emission, while the upper limits(inverted triangles) were derived specifically for the emission from the extended jet.} }
\end{figure*}

\begin{figure*}\label{fig2}
\includegraphics[angle=0,scale=0.55]{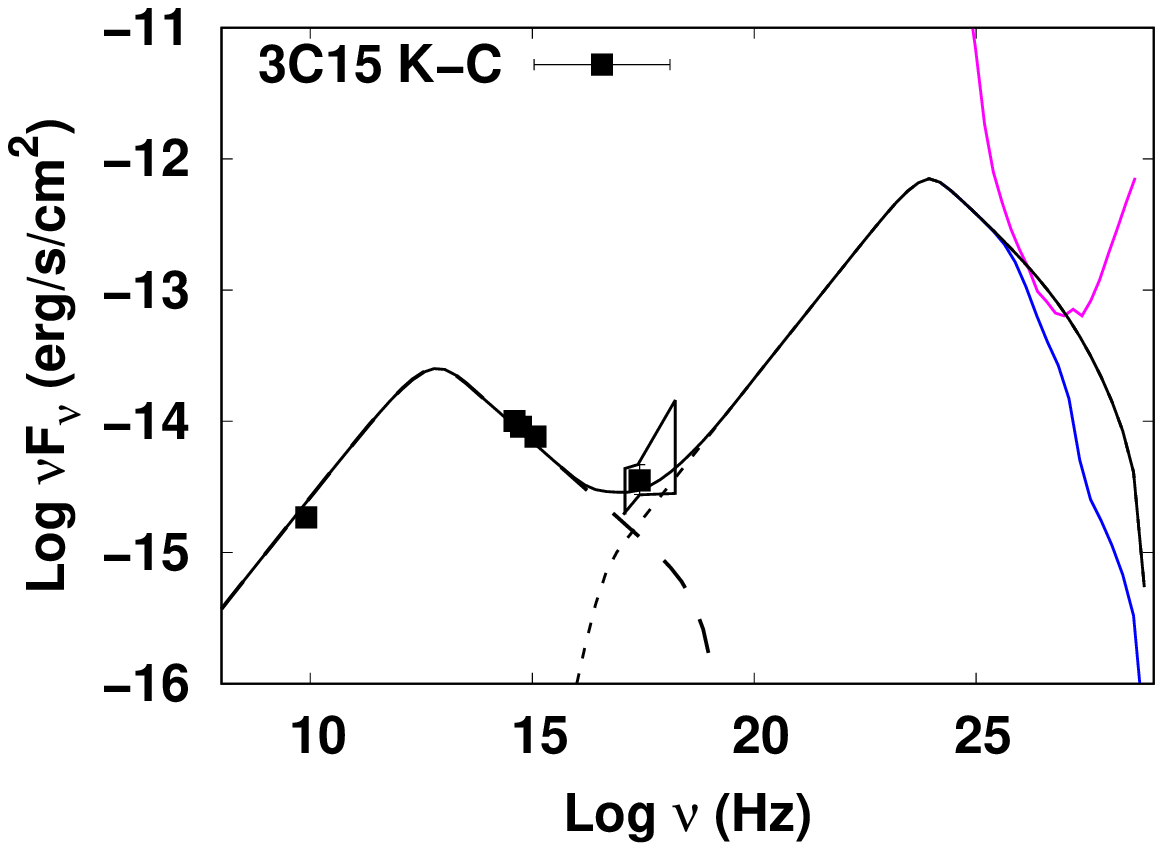}
\includegraphics[angle=0,scale=0.55]{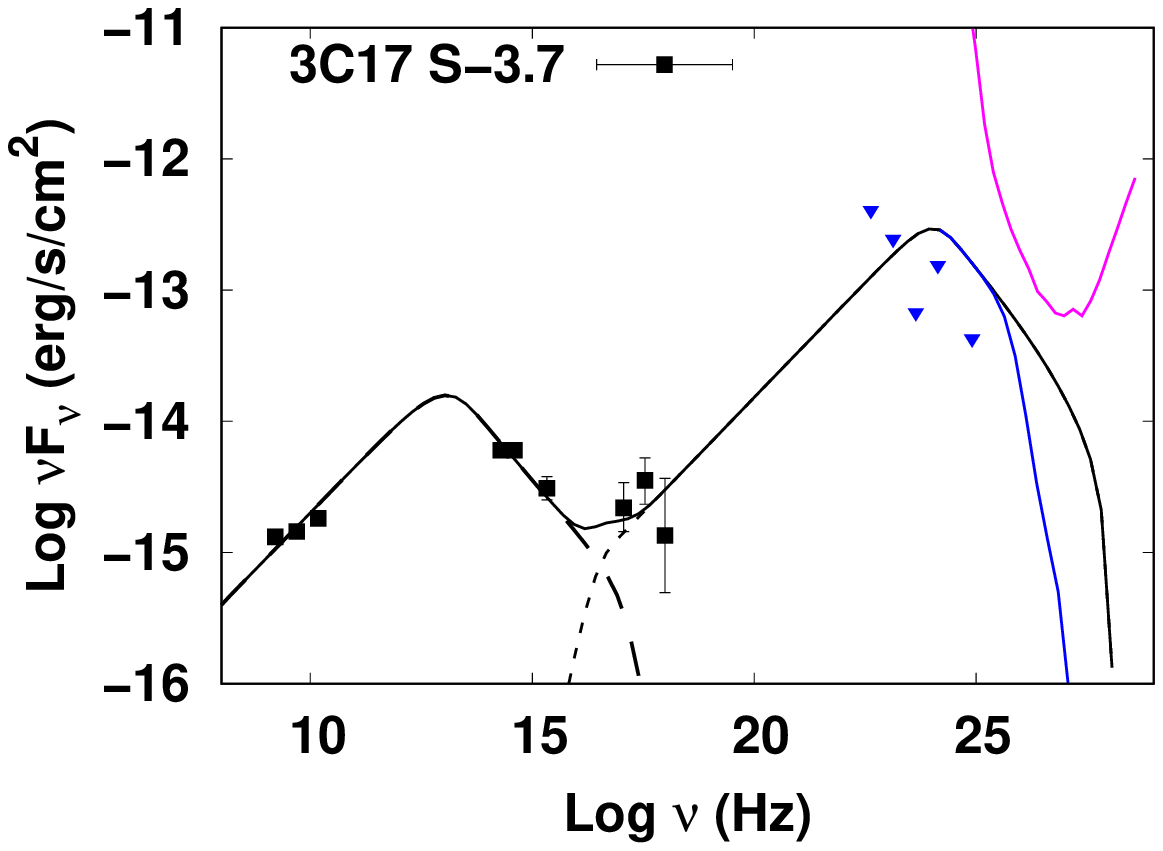}
\includegraphics[angle=0,scale=0.55]{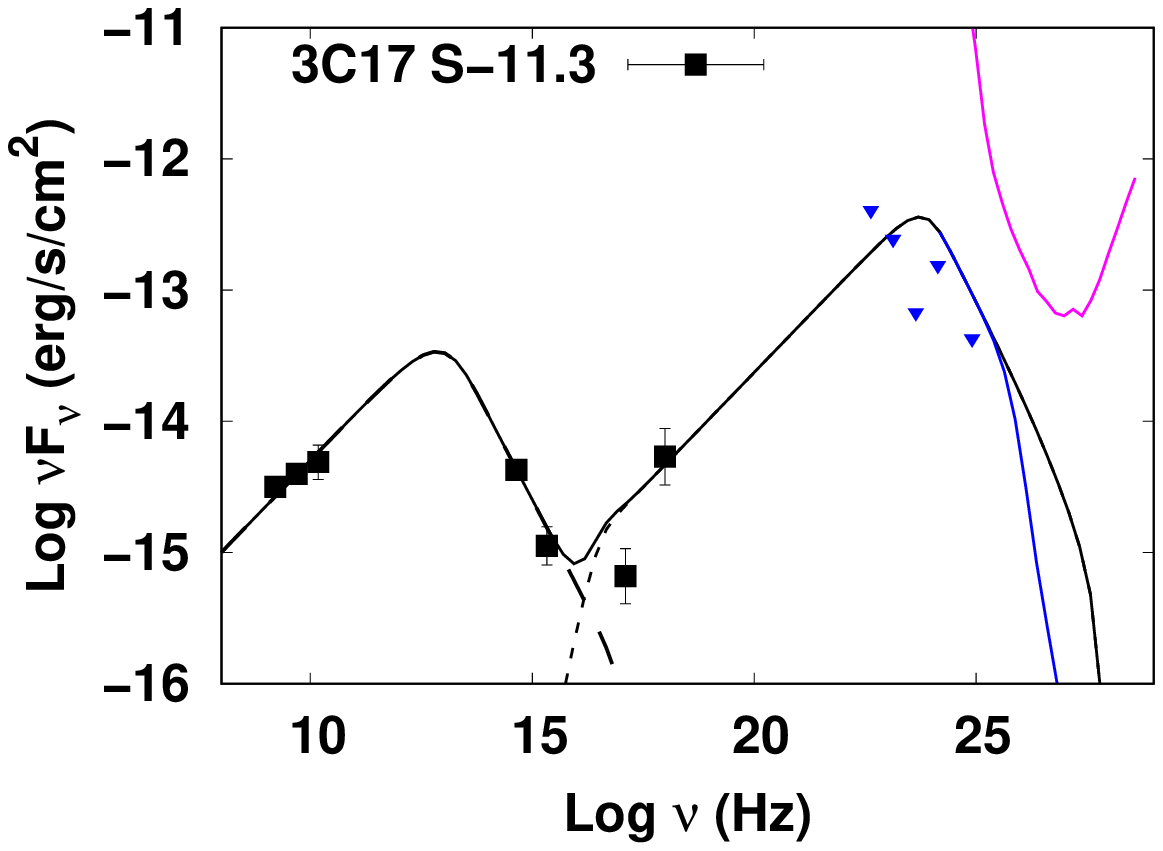}
\includegraphics[angle=0,scale=0.55]{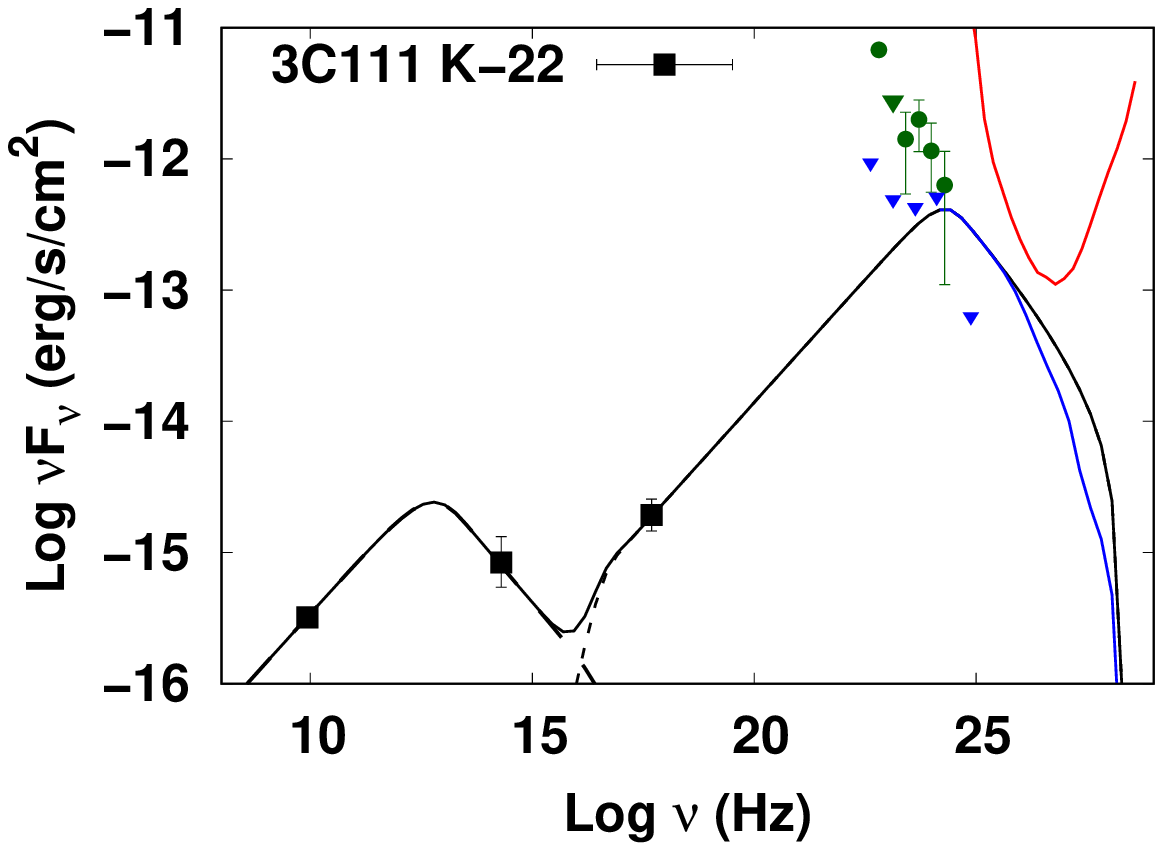}
\includegraphics[angle=0,scale=0.55]{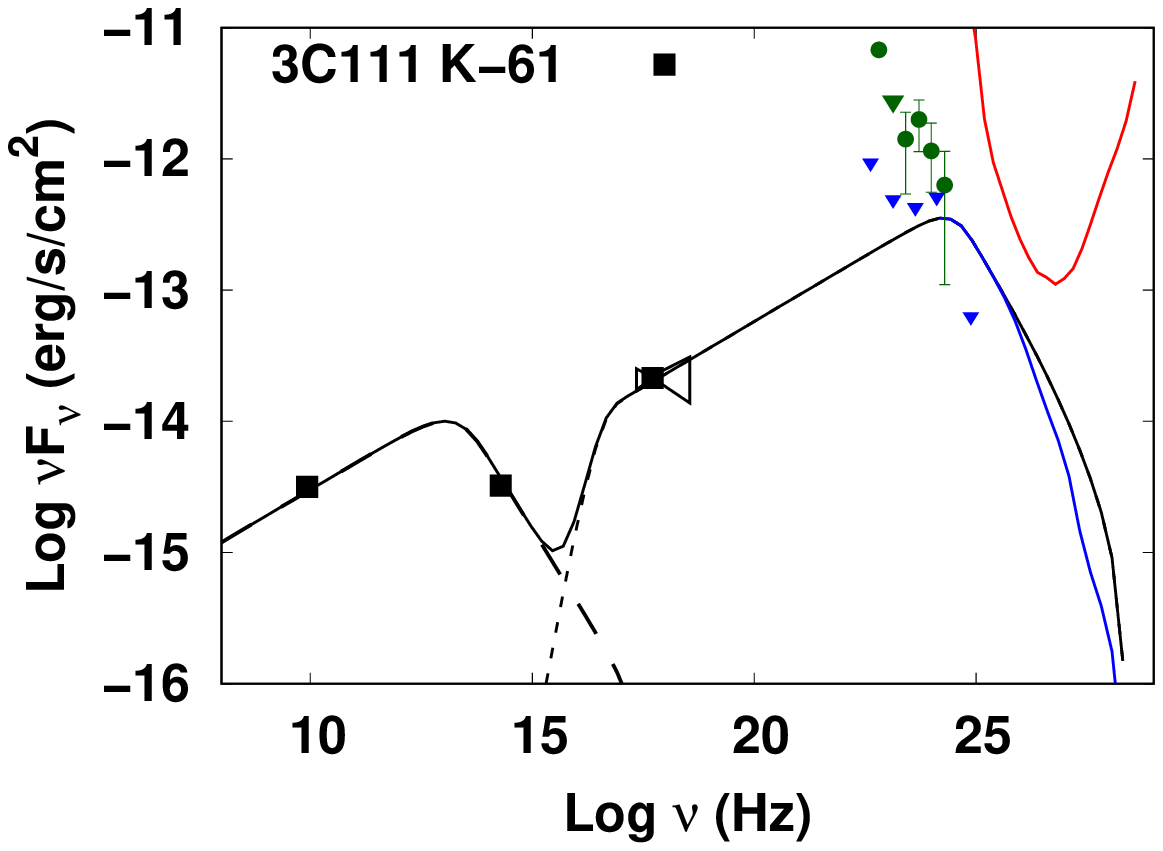}
\includegraphics[angle=0,scale=0.55]{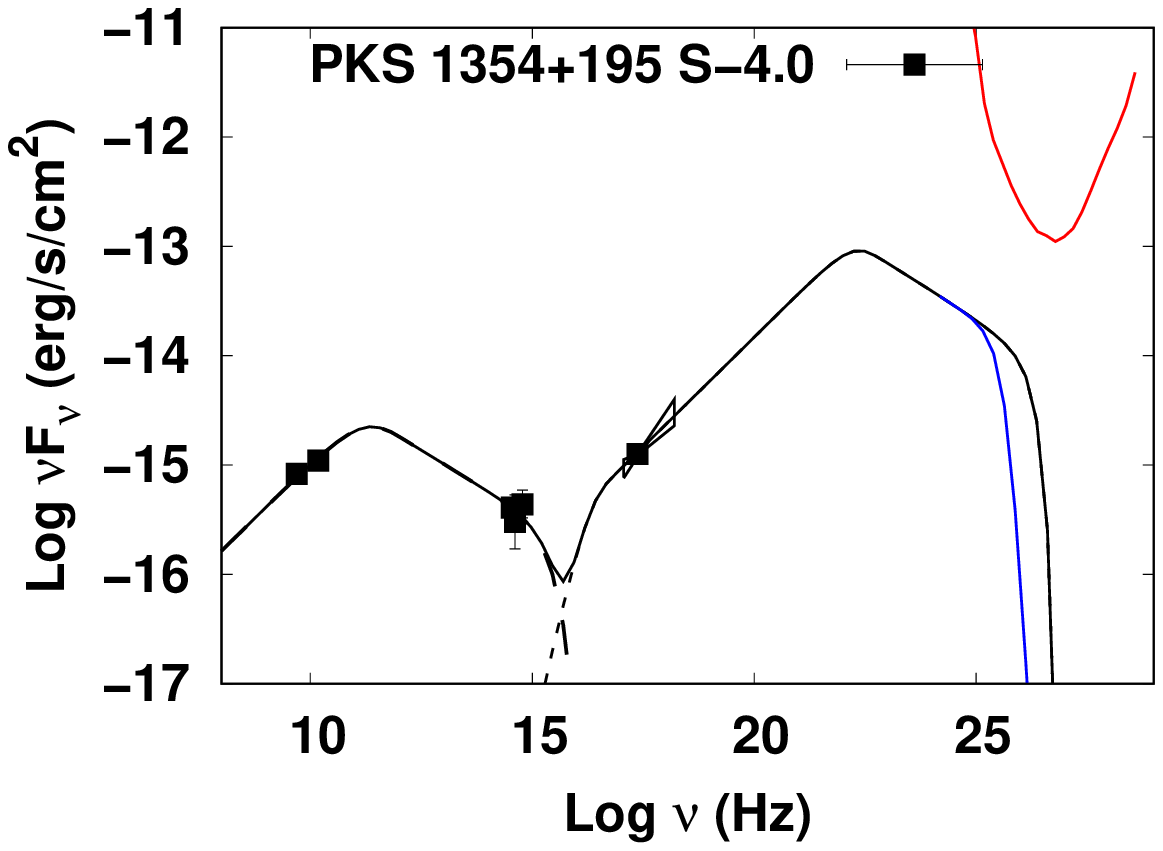}
\includegraphics[angle=0,scale=0.55]{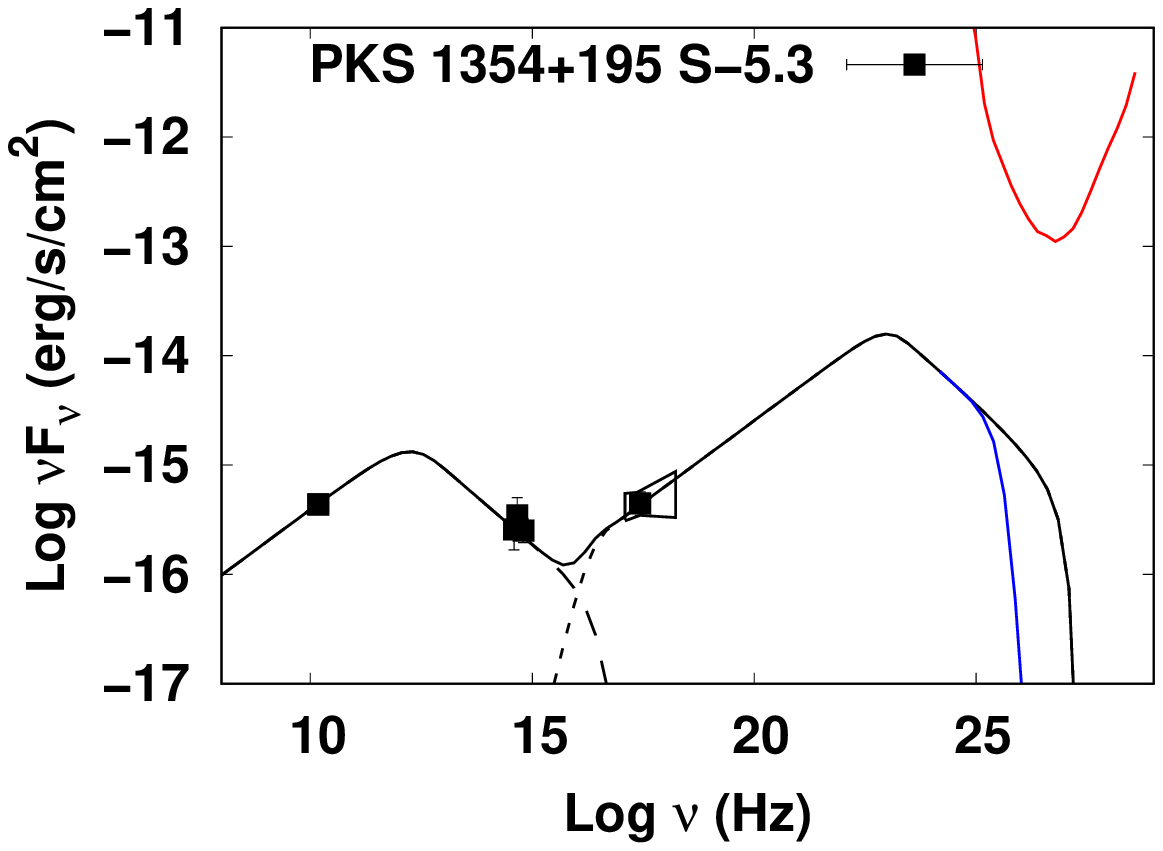}
\includegraphics[angle=0,scale=0.55]{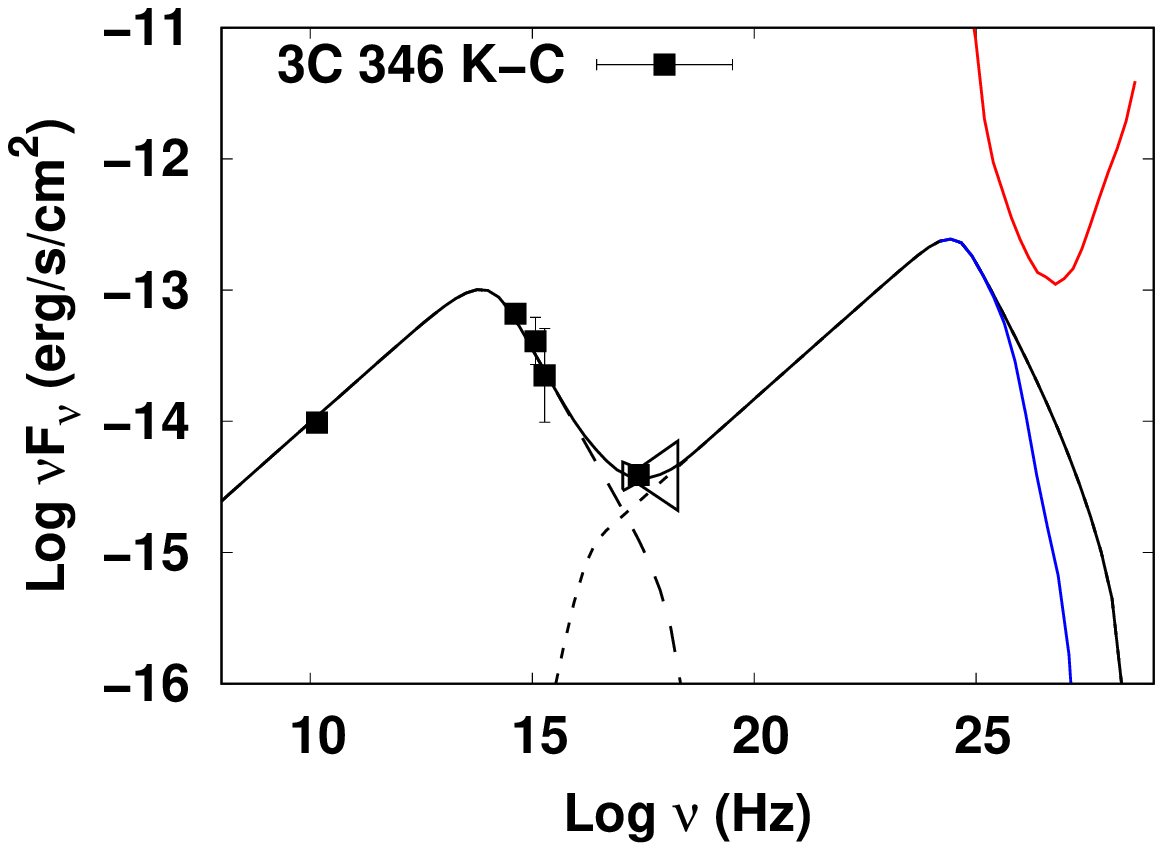}\\
\caption{Large scale knots that do not fall within the detection threshold of CTAO. Dashed line and densely-dashed line are the synchrotron and IC/CMB model curves respectively. The black solid line represents the synchtron+IC/CMB curve. Dotted lines represent the SSC model curve. Solid squares are the multiwavelength observational data of the knots.tBlue solid line is the IC/CMB model curve corrected for EBL absorption. {Red solid line and green solid line is the differential sensitivities of CTAO-Northern array Omega(50 hour) and Alpha configuration respectively}. The magenta solid curve correspond to differential sensitivities of CTAO-Southern array Omega configuration (50 hour). Inverted triangles(green) and solid circles(green) are the \emph{Fermi} upper limit values and \emph{Fermi} observations for 3C\,111 and 3C\,454.3 \citep{2017RAA....17...90X, 2015ApJ...807...51Z}. Blue inverted triangles/solid circles are the recent \emph{Fermi} upper limit values/observations \citep{2023MNRAS.518.3222B}. {Fermi-LAT points(solid circles) correspond to the total observed emission, while the upper limits(inverted triangles) were derived specifically for the emission from the extended jet.}}
\end{figure*}

\begin{figure*}

\includegraphics[angle=0,scale=0.55]{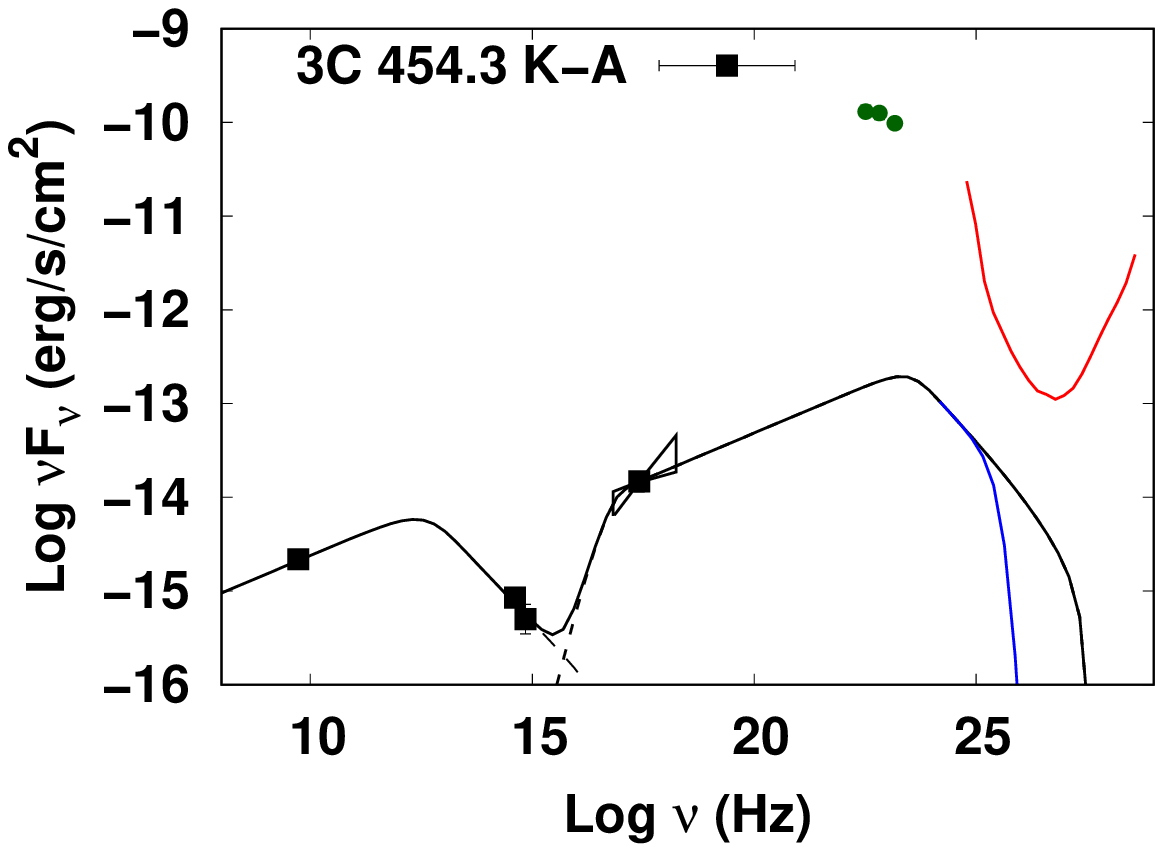}
\includegraphics[angle=0,scale=0.55]{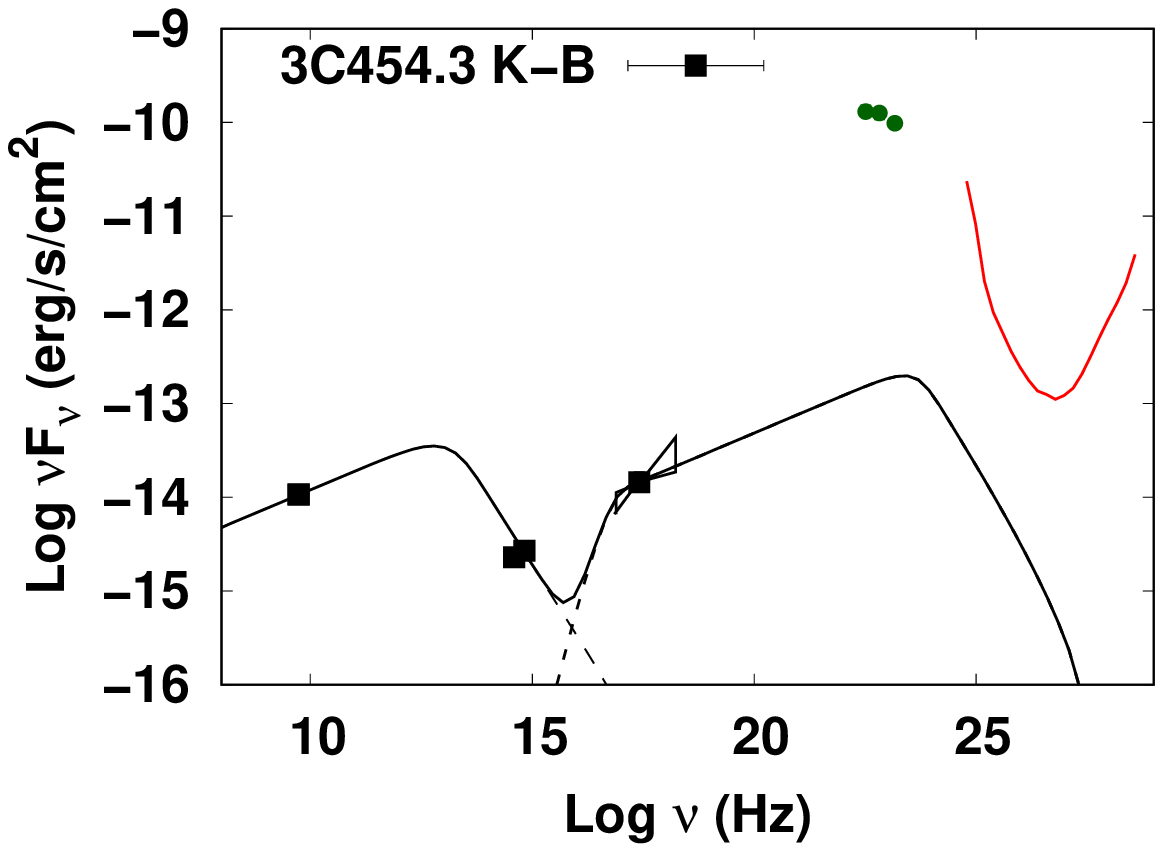}
\includegraphics[angle=0,scale=0.55]{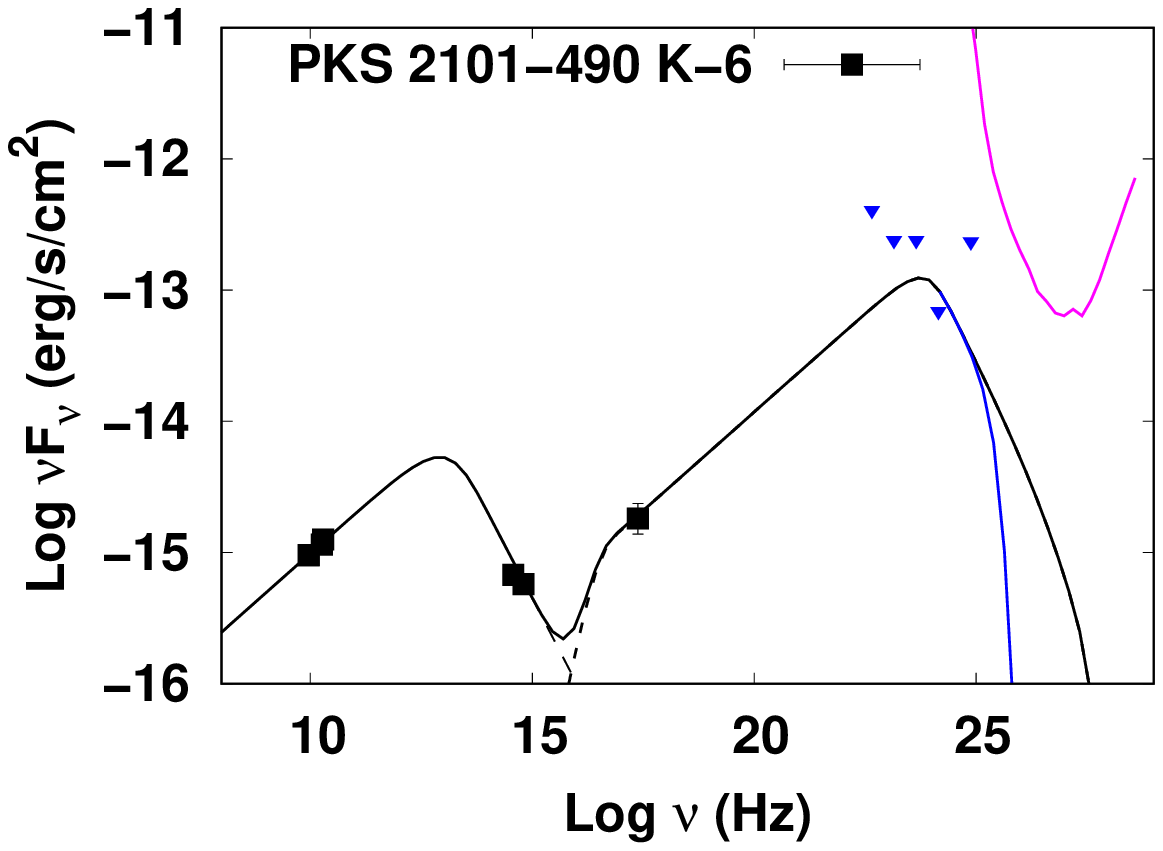}
\includegraphics[angle=0,scale=0.55]{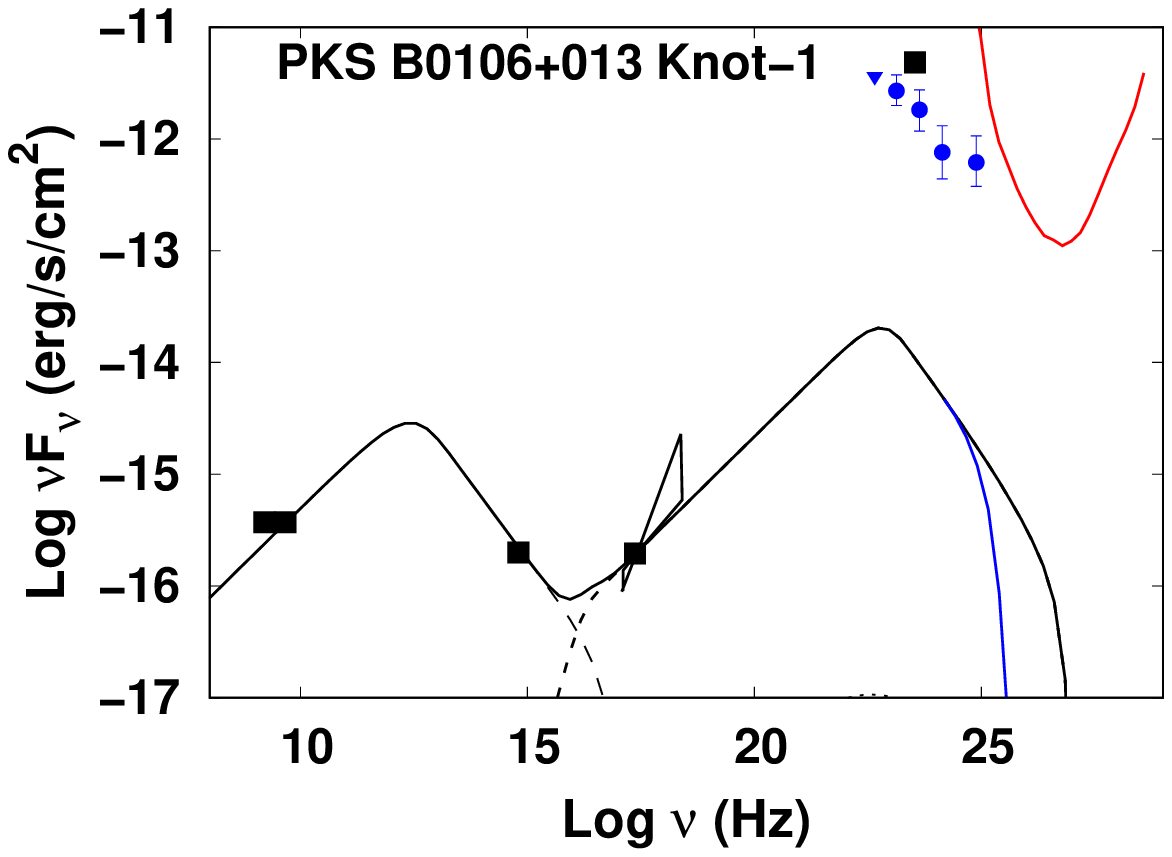}
\includegraphics[angle=0,scale=0.55]{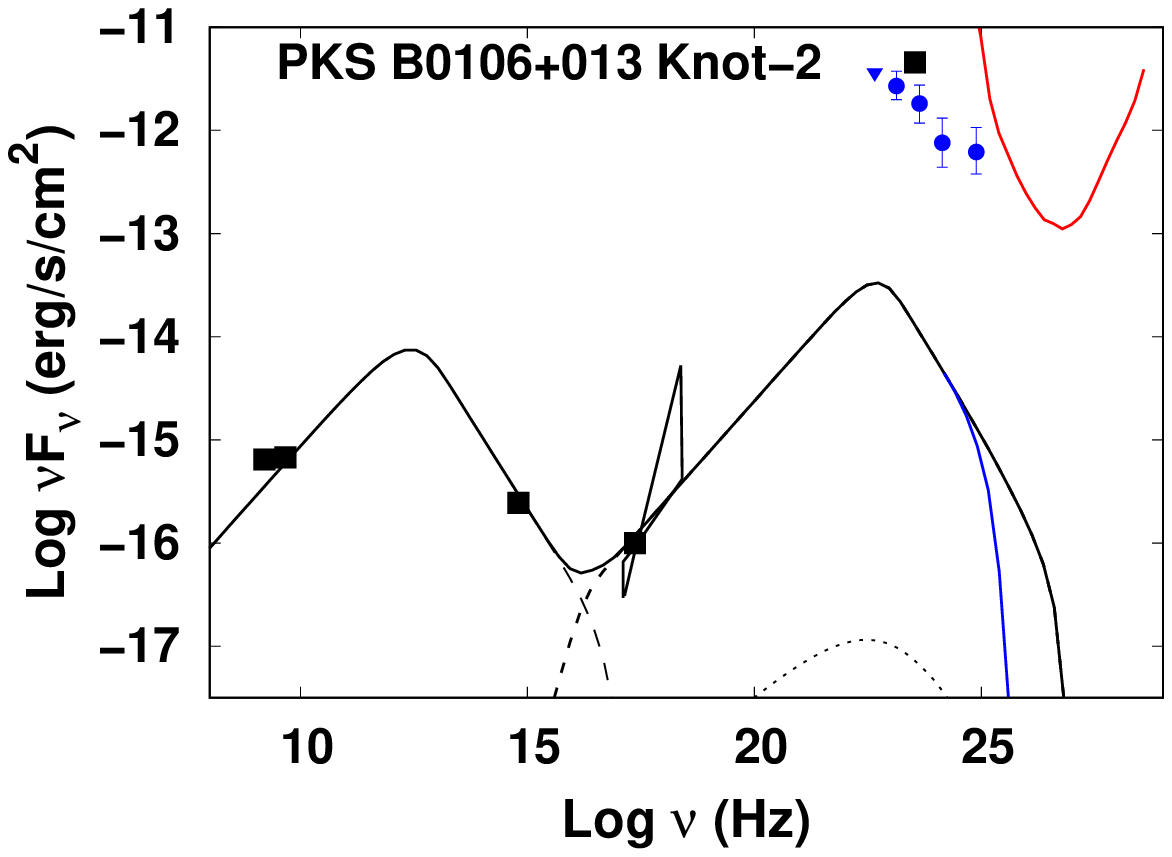}
\includegraphics[angle=0,scale=0.55]{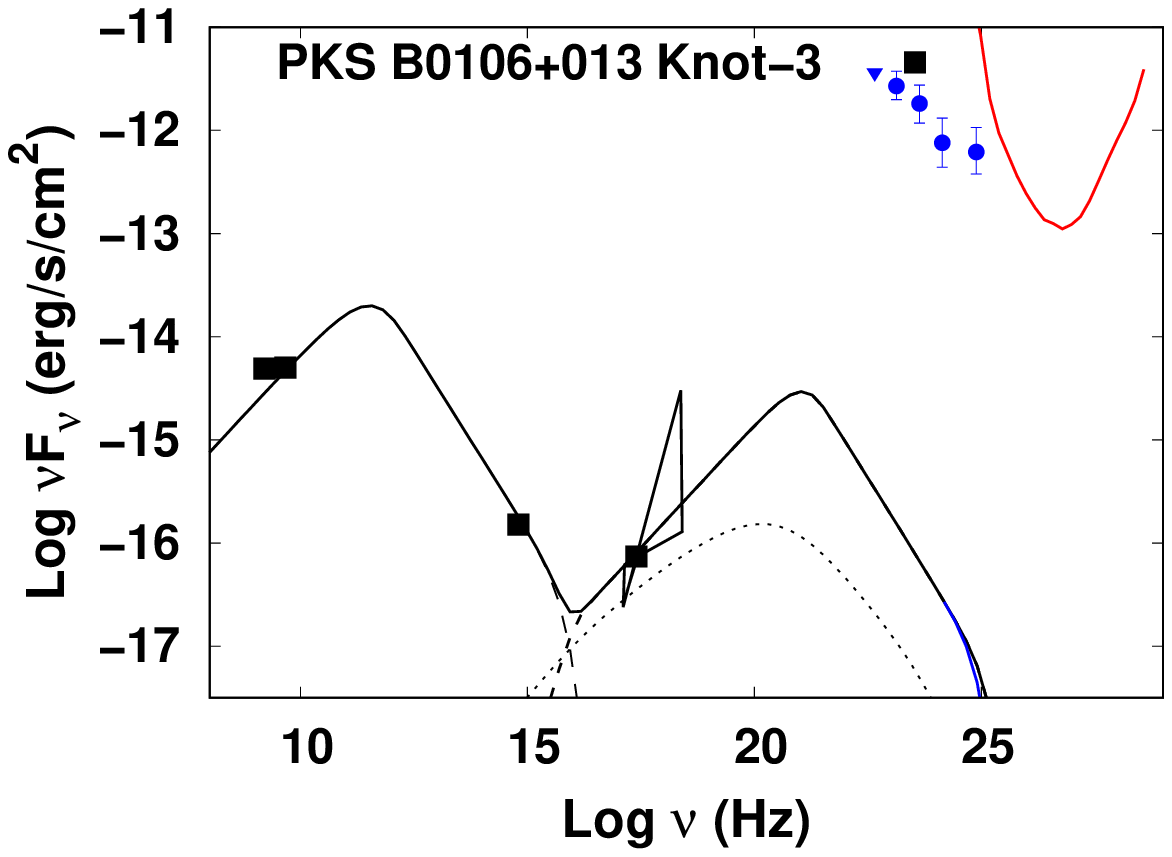}
\includegraphics[angle=0,scale=0.55]{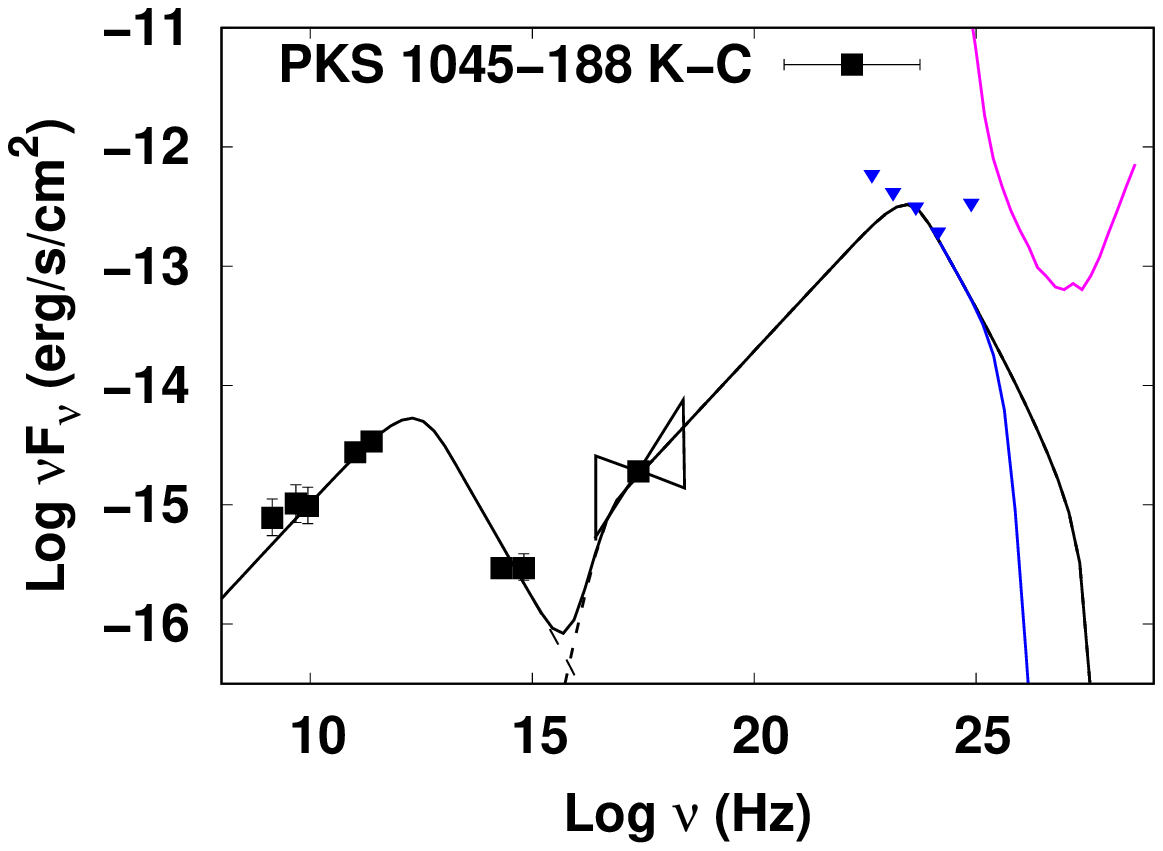}

\contcaption{}
\end{figure*}

\begin{table*}

  \centering
\begin{tabular}{|c|c|c|c|c|c|c|c|c|}
\hline
\multirow {1}{*} Source & $Knot$ & $p$      & $q$     & $R_{knot}(kpc)$ &$\Gamma$ & $\theta$ &$\gamma_b(10^6)$ & $B(10^{-6}G)$ \\ \hline
3C\,15  & K\,C                   & ${2.15}$ &${3.60}$  & ${3.0}$         &${6.0}$  &${9}$            &${0.27}$         & ${4.10}$ \\
3C\,17  & K\,S3.7                & ${2.30}$ &${3.80}$ & ${5.9}$         &${4.0}$  &${10}$         &${0.39}$         & ${5.32}$ \\
3C\,17  & K\,S11.3               & ${2.30}$ &${4.3}$  & ${4.5}$         &${6.0}$  &${10}$         &${0.29}$         &${7.78}$ \\
3C\,111 & K\,22                  & ${2.25}$ &${3.8}$  & ${4.0}$         &${8.0}$   &${9}$        &${0.44}$         &${1.53}$ \\
3C\,111 & K\,30                  & ${2.49}$ &${6.0}$  & ${5.5}$         &${4.0}$   &${9}$        &${5.92}$         &${2.34}$ \\
3C\,111 & K\,61                  & ${2.60}$ &${4.1}$  & ${4.4}$         &${6.0}$   &${9}$         &${0.52}$         &${3.45}$ \\
3C\,120 & K\,4                   & ${2.61}$ & ${5.0}$ & ${3.0}$         &${8.0}$  &${8}$         &${1.68}$         &${3.63}$ \\
3C\,120 & K\,S2                  & ${2.40}$ & ${6.0}$ & ${1.0}$         &${7.0}$   &${8}$          &${4.28}$         &${2.32}$ \\
PKS\,1354+195  & K\,S4.0         & ${2.20}$ & ${3.5}$ & ${9.0}$         &${7.5}$  &${11}$        &${0.55}$         &${6.56}$ \\
PKS\,1354+195  & K\,S5.3         & ${2.40}$ & ${3.7}$ & ${3.8}$         &${8.0}$   &${11}$          &${0.13}$         &${11.98}$ \\
3C\,346 & K\,C                   & ${2.40}$ & ${4.2}$ & ${1.8}$         &${6.0}$  &${9}$         &${0.57}$         &${15.77}$ \\
3C\,454.3 & K\,A                 & ${2.60}$ & ${4.0}$ & ${4.0}$         &${4.7}$  &${4.5}$         &${0.13}$         &${14.51}$ \\
3C\,454.3 & K\,B                 & ${2.60}$ & ${4.5}$ & ${2.2}$         &${4.7}$    &${4.5}$        &${0.15}$         &${35.50}$ \\
PKS\,2101-490  & K\,K6           & ${2.40}$ & ${4.3}$ & ${4.3}$         &${6.0}$   &${10}$          &${0.29}$         &${14.63}$ \\
PKS\,B0106+013  & Knot\,1         & ${2.20}$ & ${4.1}$ & ${2.0}$         &${3.0}$   &${13}$          &${0.12}$         &${44.73}$ \\
PKS\,B0106+013  & Knot\,2         & ${2.01}$ & ${4.40}$ & ${3.0}$        &${4.0}$   &${13}$          &${0.10}$         &${60.52}$ \\
PKS\,B0106+013  & Knot\,3         & ${2.05}$ & ${4.40}$ & ${1.0}$       &${2.0}$   &${13}$         &${0.02}$         &${255.75}$ \\
PKS\,1045-188  & K\,C           & ${2.20}$ & ${4.30}$ & ${5.0}$         &${9.0}$   &${8}$         &${0.16}$         &${6.69}$ \\

\hline
\end{tabular}\caption{Fit parameters of radio-optical-X-ray spectrum. $p$ and $q$ are the power-law indices of particle distribution;  $R_{knot}$ is the radius of the knot; $\Gamma$, $\theta$ and $\gamma_b$ are the bulk Lorentz factor, viewing angle of the jet and break energy of particle spectrum respectively. B is the magnetic field in micro-Gauss unit.}  

\label{tab2}
\end{table*}

\section{Summary}
The high energy emission from the knots of kpc scale AGN jet is often interpreted as the inverse Compton up-scattering of Cosmic microwave 
background radiation by relativistic particles in the jet (IC/CMB). However, the gamma-ray upper limits derived from the long term observation
of these sources by \emph{Fermi} disfavours this interpretation. In this work, we perform a detailed multiwavelength modelling of the knots of 
the AGN jets which are detected in X-rays using synchrotron and IC/CMB processes. The source parameters deciding the broadband spectral energy
distribution are estimated using approximate analytical expression for the emissivity functions. The emission model is extrapolated to VHE energy  
and then compared with the CTAO sensitivity.  We find the VHE model flux of certain knots/jet components of two sources, 3C\,111 and 3C\,120, fall
well above the CTAO sensitivity. However, the recently updated \emph{Fermi} upper limits again disfavors IC/CMB interpretation for the high energy emission from these sources. Therefore any sources in our study, if detected in VHE, would probably favour the second population interpretation for the high energy emission rather than IC/CMB.  




\section*{Acknowledgements}

 A.A.R acknowledge Dr.Gulab Dewangan and Inter-University Centre for Astronomy and Astrophysics(IUCAA), Pune, India for the support and research facilities. A.A.R is thankful to the financial support provided by University Grants Commission (UGC), Govt. of India. {A.A.R is also grateful to the unknown reviewer for the valuable comments}. 

\section*{Data Availability}

The data and the codes used in this work will be shared on the
reasonable request to the corresponding author Amal A. Rahman (email:amalar.amal@gmail.com)



\bibliographystyle{mnras}
\bibliography{ref} 






\bsp	
\label{lastpage}

\end{document}